\documentclass[journal]{IEEEtran}
\usepackage{enumitem}

\usepackage{cite}

\usepackage{graphicx}
\usepackage{subcaption}
\usepackage{amsmath}
\usepackage{amssymb}
\usepackage{cancel}
\usepackage{wrapfig}

\usepackage{url}

\usepackage{hyperref}

\usepackage[dvipsnames]{xcolor}

\usepackage{booktabs}
\usepackage{multirow}
\usepackage{makecell}

\usepackage{notation}

\usepackage{style}

\usepackage[colorinlistoftodos]{todonotes}

\begin{document}
%
\title{LoopTree: Exploring the Fused-layer Dataflow Accelerator Design Space}

\author{Michael~Gilbert, Yannan Nellie Wu, Joel S. Emer,~\IEEEmembership{Fellow,~IEEE,} and Vivienne Sze,~\IEEEmembership{Senior Member,~IEEE,}%
\thanks{Michael Gilbert and Vivienne Sze are with the Department of Electrical Engineering and Computer Science, School of Engineering, Massachusetts Institute of Technology, Cambridge, MA 02139, United States (e-mail: gilbertm@mit.edu, sze@mit.edu).}%
\thanks{Joel S. Emer is with the Department of Electrical Engineering and Computer Science, School of Engineering, Massachusetts Institute of Technology, Cambridge, MA 02139, United States and NVIDIA, Santa Clara, CA 95051, United States (e-mail: emer@csail.mit.edu).}%
\thanks{Yannan Nellie Wu is with Google, Mountain View, CA 94043, United States (e-mail: nellieywu@gmail.com). She completed this work while at Massachusetts Institute of Technology.}
}

%



\newcommand\copyrighttext{%
  \footnotesize \textcopyright~2024 IEEE. Personal use of this material is permitted. Permission from IEEE must be obtained for all other uses, in any current or future media, including reprinting/republishing this material for advertising or promotional purposes, creating new collective works, for resale or redistribution to servers or lists, or reuse of any copyrighted component of this work in other works.
  DOI: 10.1109/TCASAI.2024.3461716).}
\newcommand\copyrightnotice{%
\begin{tikzpicture}[remember picture,overlay]
\node[anchor=south,yshift=6pt] at (current page.south) {\fbox{\parbox{\dimexpr\textwidth-\fboxsep-\fboxrule\relax}{\copyrighttext}}};
\end{tikzpicture}%
}

\maketitle
\copyrightnotice

\begin{abstract}
Latency and energy consumption are key metrics in the performance of deep neural network (DNN) accelerators. A significant factor contributing to latency and energy is data transfers. One method to reduce transfers or data is reusing data when multiple operations use the same data. \emph{Fused-layer accelerators} reuse data across operations in different layers by retaining intermediate data in on-chip buffers, which has been shown to reduce energy consumption and latency. Moreover, the intermediate data is often tiled (\ie, broken into chunks) to reduce the on-chip buffer capacity required to reuse the data. Because on-chip buffer capacity is frequently more limited than computation units, fused-layer dataflow accelerators may also recompute certain parts of the intermediate data instead of retaining them in a buffer. Achieving efficient trade-offs between on-chip buffer capacity, off-chip transfers, and recomputation requires systematic exploration of the fused-layer dataflow design space. However, prior work only explored a subset of the design space, and more efficient designs are left unexplored. 

In this work, we propose (1) a more extensive design space that has more choices in terms of tiling, data retention, recomputation and, importantly, allows us to explore them \emph{in combination}, (2) a taxonomy to systematically specify designs, and (3) a model, LoopTree, to evaluate the latency, energy consumption, buffer capacity requirements, and off-chip transfers of designs in this design space. We validate our model against a representative set of prior architectures, achieving a worst-case 4\% error. Finally, we present case studies that show how exploring this larger space results in more efficient designs (\eg, up to a 10$\times$ buffer capacity reduction to achieve the same off-chip transfers).
\end{abstract}


%
\IEEEpeerreviewmaketitle

\section{Introduction}\label{sec:introduction}

\IEEEPARstart{D}{eep} neural networks (DNNs) are a dominant approach in various applications, such as computer vision~\cite{mobilenetv2, mobilenetv3, vggnet, alexnet}, natural language processing~\cite{nlp_from_scratch, bert, xlm, attention}, speech recognition~\cite{deepspeech}, self-driving cars~\cite{chauffeurnet}, and others. The ubiquity of DNNs, combined with the large amount of computation and data required in DNN processing, motivates the need for low-latency and energy-efficient DNN processing.

Data transfers are a significant component of energy consumption and latency in DNN processing. An effective method to reduce data transfers is by reusing data. For example, if multiple operations use the same data, data transfers from off-chip buffers can be avoided by retaining the data in an on-chip buffer to be reused for the processing of the operations. Thus, we trade off-chip transfers with the chip area allocated for on-chip buffers.

We can categorize data reuse based on the DNN structure. Structurally, DNNs are made of layers that take input feature maps (fmaps) and filters to produce output fmaps. When the operations that reuse data belong to the same layer, the reuse is an \emph{intra-layer reuse}. When the operations that reuse data belong to the different layers---specifically, the output fmap of the previous layer becomes the input fmap of the next layer (i.e, these fmaps are \emph{intermediate fmaps})---the reuse is an \emph{inter-layer reuse}.

\begin{figure}
    \centering
    \includegraphics[width=0.4895\textwidth]{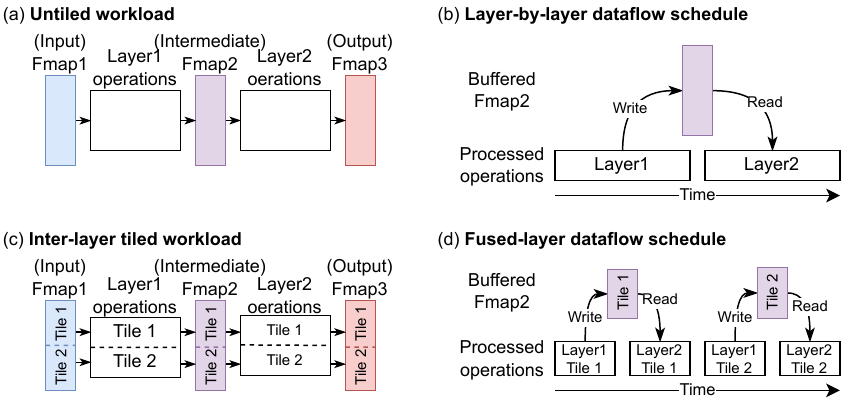}
    \caption{Comparison of layer-by-layer and fused-layer dataflows. (a) Two layers (white boxes represent operations within the layers) and three fmaps\protect\footnotemark. (b) Layer-by-layer processing produces all of Fmap2 before it is used. (c) Tiling layer operations and fmaps. (d) A fused-layer processing of Layer1 and Layer2, where only a tile of Fmap2 needs to be retained in a buffer at a time.}
    \label{fig:introduction:fusion}
\end{figure}

Many accelerators process DNNs in a \emph{layer-by-layer} fashion~\cite{eyeriss_isca, eyeriss, maeri, diannao, nvdla}, where all operations for a layer are performed before the operations for the next layer (see Fig.~\ref{fig:introduction:fusion}(a) and (b)). Because operations from a layer are scheduled close together, layer-by-layer processing is efficient at exploiting intra-layer reuse opportunities. On the other hand, reusing the intermediate fmap between layers requires retaining the entire intermediate fmap in a buffer because the entire intermediate fmap is produced before it is used in the next layer (see Fig.~\ref{fig:introduction:fusion}(b)). DNN fmaps are often large, and retaining them requires large buffers that often do not fit on-chip. More typically, layer-by-layer accelerators do not exploit inter-layer fmap reuse on-chip. Rather, intermediate fmaps are streamed to and from off-chip buffers between the processing of different layers. But this may be undesirable for two reasons. First, off-chip bandwidth is more limited than on-chip bandwidth and large volumes of off-chip transfers may cause bandwidth bottlenecks, which impacts latency. Second, off-chip transfers cost more energy than on-chip transfers.

The amount of intermediate fmap that needs to be retained to exploit inter-layer reuse can be reduced by \emph{tiling} the layers and scheduling the processing of the tiles such that an intermediate fmap tile (\ie, a chunk of the intermediate fmap) can be computed and immediately used between layers (see Fig~\ref{fig:introduction:fusion}(c) and (d)). When the intermediate fmap tile is not needed anymore, it is released from the buffer. Thus, only a tile of the intermediate fmap needs to be retained at a time. Since this tiling is applied across layers, we refer to the tiling as \emph{inter-layer tiling}. Moreover, we refer to the layers as being \emph{fused} and the accelerators employing them as \emph{fused-layer dataflow accelerators}~\cite{fusedcnn, isaac, pipelayer, tangram}\footnote{When the precision is required, we will refer to retaining entire intermediate fmaps on-chip as \emph{untiled fusion}, and fusing with inter-layer tiling as \emph{tiled fusion}.}.\footnote{The LoopTree tutorial is available at \href{https://github.com/Accelergy-Project/looptree-tutorial}{https://github.com/Accelergy-Project/looptree-tutorial}.}

In DNN accelerators, on-chip buffers often occupy a large share of the chip's area~\cite{eyeriss, eyeriss_isca, ten_lessons_tpu}. In contrast, compute units are abundant and use less energy compared to reading from buffers~\cite{eyeriss, eyeriss_isca, ten_lessons_tpu}. Fused-layer dataflows can take advantage of this fact by recomputing instead of retaining intermediate data~\cite{fusedcnn, defines, convfusion}.

\footnotetext[3]{In figures, we will show input fmaps in blue, filters in green, output fmaps in red, intermediate fmaps in purple, and operations in grayscale.}

The choices of tiling, recomputation, and retention are important design aspects that need to be explored in combination for an efficient fused-layer dataflow. However, only a subset of this design space has been explored in prior fused-layer dataflows and design space exploration (DSE) frameworks, leaving more efficient designs unexplored.

In order to precisely discuss the fused-layer dataflow design space, we briefly introduce some nomenclature (more details in Section~\ref{sec:background:einsum}). Most DNN layers can be viewed as \emph{tensor algebra operations}~\cite{extensor, edge, fusemax, teaal}. Viewed this way, the multidimensional fmaps and weights are represented by tensors, which have multiple ranks corresponding to the dimensions, \eg, channels and width.

Now, we highlight four characteristics of the fused-layer design space, and show the space of choices in prior work in Tab.~\ref{tab:introduction:comparison}.
\begin{itemize}
    \item \emph{Partitioned ranks.} Tiling is done by partitioning ranks (\eg, channels and width) in the layer. The more ranks that can be partitioned, the more choices we have to create tiles. However, most prior work only supports a limited set of ranks to partition.
    \item \emph{Recomputation.} In a given intermediate fmap, there are many ways to choose which activations (\ie, values in an fmap) are recomputed. Most prior works do not support or support only a limited set of recomputation choices.
    \item \emph{Per-intermediate-fmap recomputation.} A recomputation choice can be made for each intermediate fmap. However, prior work that supports extensive recomputation choices is limited to applying the same recomputation choice for all intermediate fmaps.
    \item \emph{Per-tensor retention.} Some prior works are limited in choosing the shape of tensor tiles that are retained. Specifically, if a rank is partitioned in the retained tile of a tensor (\ie, a filter or fmap), retained tiles of other tensors also have to partition that rank in the same way. However, being able to make this partitioning choice per tensor has been shown to significantly increase efficiency~\cite{defines, zigzag}.
\end{itemize}

\begin{table*}
\caption{Comparison of design space in prior works and this work. This work is the first to support extensive tiling, recomputation, and per-tensor retain choices.}
\label{tab:introduction:comparison}
\centering
\begin{tabular}{@{}lcccc@{}}
\toprule
    Framework & \makecell{Partitioned ranks} & \makecell{Recomputation} & \makecell[c]{Per-intermediate-fmap  recomputation} & \makecell[c]{Per-tensor retention} \\
\midrule
    TANGRAM~\cite{tangram} & \textcolor{BrickRed}{Channel} & \textcolor{BrickRed}{No} & \textcolor{BrickRed}{No} & \textcolor{OliveGreen}{Yes} \\
    DeFiNES~\cite{defines} & \textcolor{BrickRed}{Row, column} & \textcolor{OliveGreen}{All} & \textcolor{BrickRed}{No} &  \textcolor{OliveGreen}{Yes} \\
    \makecell{ConvFusion~\cite{convfusion}} & \makecell{\textcolor{BrickRed}{Row, column}, \textcolor{BrickRed}{channel}} & \makecell[c]{\textcolor{BrickRed}{Limited}\footnotemark{}} & \textcolor{BrickRed}{No} & \textcolor{OliveGreen}{Yes} \\
    Optimus~\cite{optimus} & \textcolor{BrickRed}{Row, column} & \textcolor{BrickRed}{No} & \textcolor{BrickRed}{No} & \textcolor{OliveGreen}{Yes} \\
    SET~\cite{set} & \textcolor{BrickRed}{Batch} & \textcolor{BrickRed}{No} & \textcolor{BrickRed}{No} & \textcolor{OliveGreen}{Yes} \\
    \makecell[l]{FLAT~\cite{flat}} & \makecell{\textcolor{BrickRed}{Batch, Head,} \textcolor{BrickRed}{Token}} & \textcolor{BrickRed}{No} & \textcolor{BrickRed}{No} & \textcolor{OliveGreen}{Yes} \\
    \makecell[l]{TileFlow~\cite{tileflow}} & \textcolor{OliveGreen}{Any} & \textcolor{BrickRed}{Limited}\footnotemark[4]{} & \textcolor{BrickRed}{No} & \textcolor{BrickRed}{No}  \\
    This work & \textcolor{OliveGreen}{Any} & \textcolor{OliveGreen}{All} & \textcolor{OliveGreen}{Yes} & \textcolor{OliveGreen}{Yes} \\
\bottomrule
\end{tabular}
\end{table*}

Finally, this paper presents the following contributions:

\textbf{(1) We identify a design space that supports a more extensive set of tiling, recomputation, and retention choices, and their combinations}. Our results show that exploring these choices in combination leads to more efficient designs.

\textbf{(2) We propose a taxonomy to systematically specify designs in our design space} in Section~\ref{sec:design_space}. To describe tiling in our taxonomy, we build on concepts from the Einsum notation~\cite{einsum, extensor} (reviewed in Section~\ref{sec:background:einsum}). We also discuss a fundamental relationship between data retention, reuse, refetch, and recomputation that allows us to simplify the design space while expanding the space of recomputation choices.

\textbf{(3) We present a model, LoopTree, that supports our design space and validate it}. LoopTree (discussed in Section~\ref{sec:model}), evaluates the latency, energy, amount of off-chip transfers, and required buffer capacity of a given design from our design space. In Section~\ref{sec:validation}, we validate this model against a representative set of prior architectures and show a worst-case 4\% error.

\textbf{(4) Using LoopTree, we present case studies that illustrate insights into the design of efficient fused-layer dataflows} in Section~\ref{sec:case_study}. We show that LoopTree can be used to explore the trade-off between off-chip transfers, buffer capacity, and recomputation. We will also discuss the impact of each design choice, how they interact, and how the shape of DNN layers impacts which design is more efficient.

\textbf{(5) The case studies provide insights on the impact of design choices and how to perform systematic exploration of the design space.} \Eg, based on the results and their interpretation in Section~\ref{sec:case_study:tiling_choice}, we highlight how to choose tiling and scheduling choices in order to reduce the required on-chip buffer capacity to achieve the most data reuse.

\section{Background and Motivation}\label{sec:background}
In this section, we briefly review relevant concepts in deep neural network (DNN) layers, the Einsum notation, layer fusion, and limitations in prior dataflows and DSE frameworks.

\subsection{DNN Layers}\label{sec:background:layers}
\begin{figure}[h]
    \centering
    \includegraphics[width=0.485\textwidth]{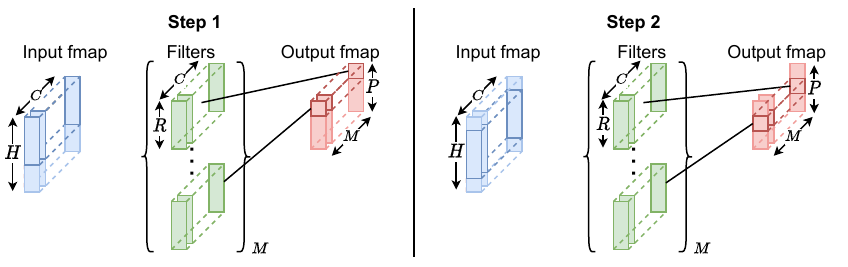}
    \caption{A 1D conv layer. All input channels ($C$) are used to generate an output fmap. Values (\ie, activations) in the output fmap column ($P$) are generated by sliding the convolution window. Different output channels ($M$) are generated with different filters.}
    \label{fig:background:conv}
\end{figure}

DNNs are composed of layers that implement certain functions. For example, Fig.~\ref{fig:background:conv} shows the computation performed by a 1-dimensional convolution (1D conv) layer. In a 1D conv layer, a window of size $R$ in the input with $C$ channels is multiplied elementwise with a filter of learned weights. The results are accumulated to create a single activation in one channel in the output fmap. The other activations in the output channel are computed by sliding the window and repeating the process. To generate $M$ output channels, $M$ filters are used, one for each output channel. We refer to the values of $H$, $R$, $C$, and $M$  ($P$ can be computed from $H$ and $R$) collectively as the layer's \emph{shape}. $H$, $R$, $C$, and $M$ themselves are referred to as \emph{ranks}. The computation performed by DNN layers can be described precisely using the \emph{extended Einsum notation}, which we discuss next.

\subsection{The Extended Einsum Notation}\label{sec:background:einsum}
The \emph{extended Einsum notation} (hereafter referred to as ``Einsum") precisely describes the computation in a layer. First, we discuss the Einsum notation. Then, we discuss how the Einsum notation helps systematize our design space.

To illustrate the Einsum notation, we look again at the 1D convolution in Fig.~\ref{fig:background:conv}. We can express this layer with a mathematical expression.
\begin{equation}\label{eq:background:1d_conv_sum}
    \text{Output}[m,p] = \sum_{c=0,r=0}^{c=C-1,r=R-1}{\text{Input}[c,p+r] \times \text{Filter}[m,c,r]}
\end{equation}
where $m \in [0,M), p \in [0,P)$.

\footnotetext[4]{Only a subset of recomputation choices are supported.}

The Einsum notation captures the same computation as Eq.~\ref{eq:background:1d_conv_sum} succinctly by allowing implicit inference of the attributes of the summation. In the Einsum notation, Output, Input, and Filter are \emph{tensors} (\ie, multi-dimensional arrays). Each tensor has named \emph{ranks} (written in uppercase letters) and indices that iterate along ranks (written in lowercase). The shape of each tensor is specified by writing the shape of each rank (\ie, the range of legal index values) after an equal sign in the superscript\footnote{This notation is extended from the original Einsum~\cite{einsum} by Hedge et al.~\cite{extensor} to allow affine expressions in the index. Writing the shape of tensors in the superscript is our further extension.} (Tab.~\ref{tab:design_space:ranks} lists commonly used ranks). Eq.~\ref{eq:design_space:einsum} shows the Einsum equivalent to Eq.~\ref{eq:background:1d_conv_sum} if we also provide the numerical shape of each rank.

\begin{equation}\label{eq:design_space:einsum}
    \text{Output}_{m,p}^{M=4,P=6} = \text{Input}_{c,p+r}^{C=3,H=8}\text{Filter}_{m,c,r}^{M=4,C=3,R=3}
\end{equation}

\begin{table}[b]
\centering
\caption{Commonly used ranks in convolutional and transformer layers and their definitions.}
\label{tab:design_space:ranks}
\begin{tabular}{@{}cl|cl@{}}
\toprule
\multicolumn{2}{c|}{\textbf{Ranks in convolutional layers}} & \multicolumn{2}{c}{\textbf{Ranks in transformer layers}} \\
Rank & Definition & Rank & Definition\\
\midrule
$B$ & Batches                & $B$ & Batches \\
$P$ & Output height (rows)   & $H$ & Heads \\
$Q$ & Output width (columns) & $M$ & Tokens \\
$M$ & Output channels        & $E$ & Output embedding \\
$C$ & Input channels         & $D$ & Input embedding \\
$H$ & Input height (rows)    \\
$W$ & Input width (columns)  \\
$R$ & Kernel height \\
$S$ & Kernel width \\
\bottomrule
\end{tabular}

\end{table}

Note that we have explicitly named the input height rank $H$ in Eq.~\ref{eq:design_space:einsum} for completeness, whereas Eq.~\ref{eq:background:1d_conv_sum} does not. Frequently, the superscripts are easily inferred and omitted for brevity.

Now, we discuss how the Einsum notation helps us describe tiling systematically. The tiling of a tensor can be described as the partitioning of ranks (\eg, splitting the $M=4$ output channels into $M_1=2$ tiles with $M_0=2$ channels each). Most prior works assume a preset selection of ranks that can be partitioned (see Tab.~\ref{tab:introduction:comparison}). As we discuss later, in our design space the user specifies layers as Einsums, and any of the ranks can be partitioned.

Partitioning different ranks results in tiles of operations that reuse tiles of Input, Output, and Filter (\ie, tiles of the data) differently (see Tab.~\ref{tab:background:tiled_rank_reuse}). We can arrive at the data reuse pattern by inspecting the Einsum. For example, say we partition rank $P$ in the Einsum in Eq.~\ref{eq:design_space:einsum} (see first row in Tab.~\ref{tab:background:tiled_rank_reuse}). Then, we define tiles in Input, Output, and Filter to be the subsets of the tensors accessed by the operation tiles. When data tiles overlap, the data in the overlapping regions are reused. The rank $P$ appears as part of an affine expression $p+r$ in Input, thus partitioning $P$ results in Input tiles that form sliding windows. The activations in the partially overlapping regions are reused in a pattern we refer to as convolutional reuse. Because $P$ appears on its own in Output, the Output tiles do not overlap (\ie, no reuse). Finally, because $P$ does not appear in Filter, partitioning $P$ results in Filter tiles that are the entirety of Filter (\ie, full reuse).

\begin{table}[b]
\caption{Comparison of data reuse patterns in input fmap, output fmap, and filter tensors when partitioning different ranks in a CNN. ``Conv." refers to convolutional reuse where tiles partially overlap, as opposed to a ``Full" reuse where tiles overlap fully.}
\label{tab:background:tiled_rank_reuse}
\centering
\begin{tabular}{@{}lcccc@{}}
\toprule
    \multirow{1}{*}{Partitioned ranks} & \multicolumn{3}{c}{Data reuse in} \\
    \cmidrule{2-4}
    & \makecell{Input} & \makecell{Output} & Filter \\
\midrule
    Row, column ($P$, $Q$) & Conv. & None & Full \\
    Input channel ($C$) & Full & Full & None\\
    Tokens ($M$) & None & None & Full \\
\bottomrule
\end{tabular}

\end{table}

\subsection{Tiling and Retention-recomputation Choices in Fused-layer Dataflows}\label{sec:background:fused_layer}

In this section, we provide a brief background on tiling, recomputation, and their interaction in fused-layer dataflows.

\begin{figure}[t]
    \centering
    \includegraphics[width=0.49\textwidth]{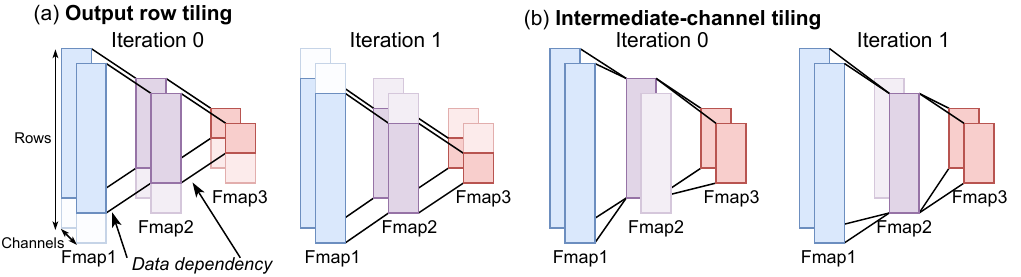}
    \caption{Examples of different tiling choices. (a) Tiles (darker shade) of Fmap1, Fmap2, and Fmap3 in iterations 0 and 1 in an output row ($P2$ rank) tiling. Note that Fmap2 tiles overlap using this tiling. (b) Tiles (darker shade) of Fmap1, Fmap2, and Fmap3 in iterations 0 and 1 in an intermediate-channel ($C2$ rank) tiling. Note that Fmap2 tiles do \emph{not} overlap using this tiling.}
    \label{fig:background:tiling}
\end{figure}

Fused-layer dataflows tile layers in order to retain only tiles of intermediate fmaps, thus reducing required on-chip buffer capacity, while still reusing the intermediate fmaps. Fig.~\ref{fig:background:tiling}(a) shows an example of tiling by partitioning the output row of the second layer (\ie, rank $P2$,  as in rank $P$ of layer 2).

Fused-layer dataflows can also reduce required on-chip buffer capacity through recomputation. For example, note that tiles in iterations 0 and 1 in Fig.~\ref{fig:background:tiling}(a) overlap. The overlap contains activations in Fmap2 that are computed and used in iterations 0 and 1. At iteration 0, we have two choices: (1) retain the common activations in a buffer to reuse in iteration 1, or (2) do not retain them to save buffer capacity but recompute them later. This \emph{retention-recomputation} choice trades off the buffer capacity required for the intermediate fmap tile with extra computation.

Note that the tiling choice determines the space of retention-recomputation choices. For example, in Fig.~\ref{fig:background:tiling}(b), tiles of Fmap2 are created by partitioning channels of Fmap2. Because the tiles of Fmap2 in different iterations do not overlap, this tiling choice results in no retention-recomputation choice. This relationship between tiling and retention-recomputation choices can only be analyzed by exploring them in combination.

\subsection{Limitations in Prior Fused-layer Dataflows}\label{sec:background:limitations}
Prior fused-layer dataflows and design space exploration (DSE) frameworks have explored only a limited subset of the fused-layer dataflow design space, leaving more efficient designs unexplored (see Tab.~\ref{tab:introduction:comparison}). We list these limitations and discuss why addressing them leads to more efficient designs.

\begin{figure}[h]
    \centering
    \includegraphics[width=0.4\textwidth]{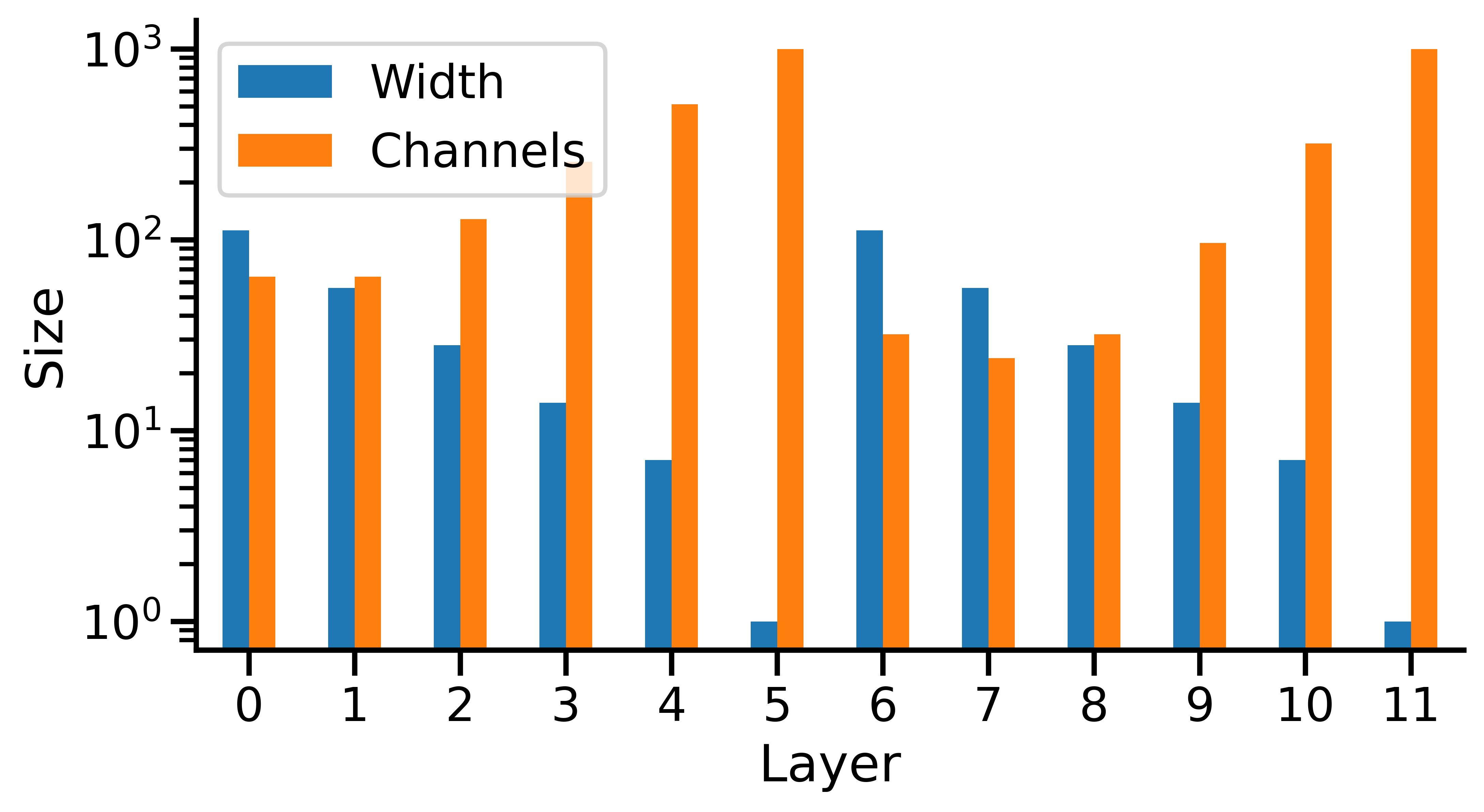}
    \caption{The width (height is the same as width) and channels of layers in ResNet-18\cite{resnet} (layers 1-5) and MobileNetv2~\cite{mobilenetv2} (layers 6-11) vary by orders of magnitude.}
    \label{fig:background:shape}
\end{figure}

\textbf{Limitation 1: tiling choices}. As shown in prior work, tiling is needed to efficiently exploit reuse~\cite{eyeriss_isca, timeloop, set, tileflow, flat}. Tiling is done by partitioning ranks and, as shown in Tab.~\ref{tab:background:tiled_rank_reuse}, the choice of partitioned rank impacts the reuse of all tensors. Furthermore, the shape of the layer, which is diverse in DNNs (see Fig.~\ref{fig:background:shape}), determines the amount of reuse. Combined, this might make a ``Full" reuse of a smaller fmap less than a partial ``Conv." reuse of a larger fmap. In Section~\ref{sec:case_study}, we show that different tiling choices may lead to $10\times$ larger required buffer capacity and that no single tiling choice is universally optimal for every layer shape. However, among prior work, only TileFlow~\cite{tileflow} supports an extensive set of partitioned rank choices.

\textbf{Limitation 2: support for recomputation.} As we discussed in Section~\ref{sec:background:fused_layer}, fused-layer dataflows can trade off buffer capacity with recomputation. We also discussed that the tiling choice impacts recomputation choices and these choices need to be explored in combination. In Section~\ref{sec:case_study}, we show that recomputation may reduce buffer capacity by $2\times$ while adding only 10\% more computation. However, prior models either support a limited choice of partitioned ranks or do not support recomputation (see Tab.~\ref{tab:introduction:comparison}).

\textbf{Limitation 3: support for per-intermediate-fmap recomputation choices.} Conceptually, retention-recomputation choices can be made per-intermediate-fmap. As we will discuss in Section~\ref{sec:case_study:per_fmap_choice}, being able to make the choice per-intermediate-fmap matters. However, no prior work supports per-intermediate-fmap recomputation choices, and all intermediate fmaps must use the same choice.

\textbf{Limitation 4: support for per-tensor retention choices}. As discussed before, given a particular tiling, the tensors in our layers may be reused differently (see Tab.~\ref{tab:background:tiled_rank_reuse}). It has been shown that making retain choices per tensor, adapting to the particular reuse of each tensor, can lead to more efficient dataflows~\cite{zigzag}. In Section~\ref{sec:case_study}, we show that per-tensor retention may reduce buffer capacity by $10\times$. However, none of the prior work that addresses limitations 1 and 2 supports per-tensor retain choices.

\section{A More Extensive and Systematic Fused-layer Dataflow Design Space}\label{sec:design_space}

In this section, we discuss a design space that addresses the limitations in Section~\ref{sec:background:limitations}. Formally, we express the design space by describing \emph{mappings}, which is the way operations and data are tiled and scheduled to buffers and compute units~\cite{eyeriss_isca, timeloop}. In describing the mappings below, we assume that the user has defined a set of layers to fuse, referred to as a \emph{fusion set}, and an architecture expressed as a set of buffers and compute units. Methods for finding the optimal fusion sets have been explored extensively in prior work (see Section~\ref{sec:related_works}).

To create a mapping, the user makes several mapping choices (or, ``choices" for short). We list these choices in Tab.~\ref{tab:design_space:design_space}. In this paper, we focus on the subset of choices specific to the design of fused-layer dataflows, which we refer to as \emph{inter-layer} choices (see Table~\ref{tab:design_space:design_space}). Most of the following subsections explain these inter-layer choices. However, we also need to describe how the tile of each layer is processed by specifying the \emph{intra-layer} mapping choices. Intra-layer mapping choices are not the focus of this paper, but poor intra-layer mapping choices will negatively impact the latency and energy of the accelerator~\cite{timeloop, zigzag, maestro, sparseloop}. Thus, LoopTree supports them. We discuss intra-layer mapping choices in Section~\ref{sec:design_space:intra_layer}.

\renewcommand{\arraystretch}{1.3}
\begin{table}[h]
\centering
\caption{Mapping choices in LoopTree.}
\label{tab:design_space:design_space}
\begin{tabular}{@{}lp{5cm}@{}}
\toprule
Mapping choices & \makecell[c]{Space of choices} \\
\midrule
Partitioned ranks                       & A subset of ranks from the last layer \\
Tile shape                              & An integer for each partitioned rank \\
\makecell[l]{Tile processing schedule}  & A permutation of the partitioned ranks \\
Retain-recompute                        & One of the partitioned ranks for each intermediate fmap \\
Retain-refetch                          & One of the partitioned ranks for each non-intermediate-fmap tensor \\
Parallelism                             & ``Sequential" or ``Pipeline" \\
\bottomrule
\end{tabular}
\end{table}
\renewcommand{\arraystretch}{1}

\subsection{Tiling by Partitioning Ranks}\label{sec:design_space:partitioning}
\begin{figure}[h]
    \centering
    \includegraphics[width=0.485\textwidth]{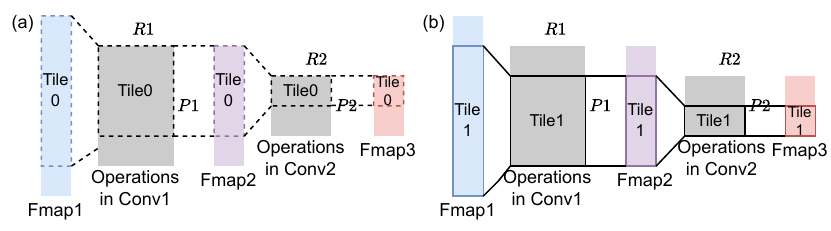}
    \caption{Partitioning rank $P2$ in Conv2 to create two Conv2 tiles, Tile0 and Tile1. (a) Given the the operation Conv2 Tile0, other data and operation tiles can be inferred. (b) The same is true for Tile1.}
    \label{fig:design_space:partitioning_P2}
\end{figure}

\begin{figure}[h]
    \centering
    \includegraphics[width=0.48\textwidth]{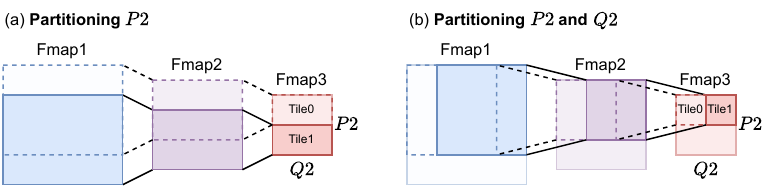}
    \caption{Examples of partitioning ranks in a 2D conv. (a)~Partitioning just $P2$. (b) Partitioning $P2$ and $Q2$.}
    \label{fig:design_space:partitioning_multiple_ranks}
\end{figure}

Fused-layer dataflows employ inter-layer tiling to reduce the buffer capacity required to exploit inter-layer reuse. In LoopTree, we define tiles of the operation space (\ie, the set of operations) of the last layer in the fusion set by partitioning its ranks. Tiles of the data can be determined from the operation tiles via data dependencies (see Fig.~\ref{fig:design_space:partitioning_P2}). Furthermore, we only need to specify the tiling of the last layer because the tiling of all the data and operations of earlier layers can be inferred through data dependencies (see Section~\ref{sec:model:analyzing_tile_shapes} for more detail). In general, we can partition any number of ranks to define our tiles (Fig.~\ref{fig:design_space:partitioning_multiple_ranks} shows an example). We can also partition the same rank multiple times, which may be useful in architectures with multiple buffer levels. For example, we can partition $P2$ to create tiles such that it fits in a buffer, then partition $P2$ again to get smaller tiles to fit in a lower-level buffer. Note that we have only discussed how to define tiles, while Section~\ref{sec:design_space:retention} discusses how to specify which tiles are retained in buffers.

To finish defining our tiles, we specify the shape of the tile, which is the length of the tile along every rank. The user only needs to specify the length of the tile along partitioned ranks because the tile is assumed to extend to the full length of unpartitioned ranks (\eg, in Fig.~\ref{fig:design_space:partitioning_P2}, only the shape of the tile along $P2$ needs to be specified because the tile extends fully along unpartitioned ranks, such as $R2$). The shape of the tile is commonly chosen such that the tile fits in a buffer.

\subsection{Tile Processing Schedule}\label{sec:design_space:schedule}
After defining the operation tiles, we can specify the scheduling of the operation tiles, which is important because it determines the order in which data is accessed and thus how long we must retain data to achieve a certain amount of reuse. Generally, retaining data for longer requires larger buffers.

Our scheduling follows the constraint that the output from one layer is immediately consumed by the next layer. \Eg, in Fig.~\ref{fig:design_space:partitioning_P2}, if we schedule Conv1 Tile0 before Conv1 Tile1, then we produce Fmap2 Tile0 before Fmap2 Tile1 (as in Fig.~\ref{fig:design_space:partitioning_P2}(a)). Thus, we schedule the operations in Conv2 Tile0, which consumes Fmap2 Tile0, before Conv2 Tile1, which consumes Fmap2 Tile1. 

In LoopTree, we schedule tile processing by specifying the ranks we want to iterate over, similar to writing loops in a loop nest from the outermost to the innermost. For example, if we follow the $P2,Q2$ schedule to process the tiles in Fig.~\ref{fig:design_space:partitioning_multiple_ranks}, we will process Tile0, Tile1, then the row underneath. In other words, we iterate across $Q2$ (within a row) first before iterating across $P2$ (across rows). (This schedule is also known as the \emph{raster scan} pattern.) We can also specify $Q2,P2$ where we iterate across $P2$ (within a column) first before going across $Q2$ (across columns).

\subsection{Parallelism}\label{sec:design_space:parallelism}
\begin{figure}[h]
    \centering
    \includegraphics[width=0.48\textwidth]{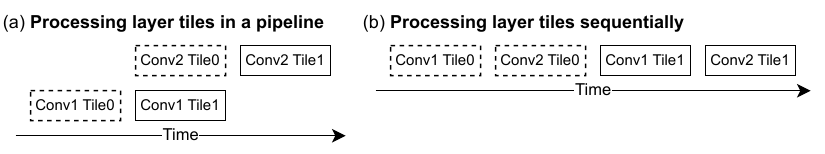}
    \caption{Parallelism choices: (a) Scheduling tiles across layers in a pipeline (\ie, in parallel); and, (b) Scheduling tiles across layers sequentially.}
    \label{fig:design_space:inter_layer_parallelism}
\end{figure}

Within each layer, we process tiles in the order we have specified in Section~\ref{sec:design_space:schedule}. The scheduling of operation tiles in different layers also needs to follow data dependencies (\eg, in Fig.~\ref{fig:design_space:partitioning_P2}, Conv2 Tile0 has to be processed after Conv1 Tile0). However, we have the freedom to choose the relative timing of subsequent tiles of different layers (\eg, in Fig.~\ref{fig:design_space:partitioning_P2}, Conv1 Tile1 and Conv2 Tile0). We can arrange these tiles to be processed sequentially or in a pipeline (see Fig.~\ref{fig:design_space:inter_layer_parallelism}).

The parallelism choice does not address the limitations of prior work mentioned in Section~\ref{sec:background:limitations}. However, we support it to be comprehensive such that we can model prior accelerators (see Section~\ref{sec:validation}).

\subsection{Data Retention}\label{sec:design_space:retention}
To achieve on-chip reuse of data, we need to specify which tiles to retain in on-chip buffers and for how long. This choice will determine the required on-chip buffer capacity and the amount of reuse. Here, we will discuss how retention choices are specified in LoopTree, assuming a two-level memory hierarchy of on-chip and off-chip buffers to simplify our discussion.

However, we first observe that although prior work has proposed recomputation as a separate design choice~\cite{fusedcnn, defines}, recomputation can be seen as a consequence of our processing schedule and retention choice. Specifically, if we specify a processing schedule and a retention choice such that an operation needs to access an intermediate fmap activation that is not retained in on-chip nor off-chip buffers, then we have to recompute that activation. Note the resemblance with non-intermediate fmap tensors where data that is not retained in on-chip buffers need to be \emph{refetched} from off-chip buffers (note that non-intermediate fmap tensors need to be backed in off-chip buffers). This observation allows us to simplify the design space without restricting the design space: we can specify retention-recomputation choices of intermediate fmaps and retention-refetch choices of non-intermediate fmap tensors using the same representation, which we discuss next.

Section~\ref{sec:design_space:partitioning} discussed how ranks are partitioned to create operation tiles and data tiles are calculated from operation tiles using data dependencies. In LoopTree, we make a retention choice for each tensor by choosing the last rank partitioned to form the retained tile, which can be one or none of the partitioned ranks, and a buffer in the architecture that retains the data. To illustrate, consider again the tiles formed by partitioning $P2$ and $Q2$ in Fig.~\ref{fig:design_space:partitioning_multiple_ranks}. We can retain the entirety of Fmap2, a tile of Fmap2 formed by partitioning the $P2$ rank, or a tile of Fmap2 formed by partitioning the $P2$ and $Q2$ ranks.

\begin{figure}[b]
    \centering
    \includegraphics[width=0.49\textwidth]{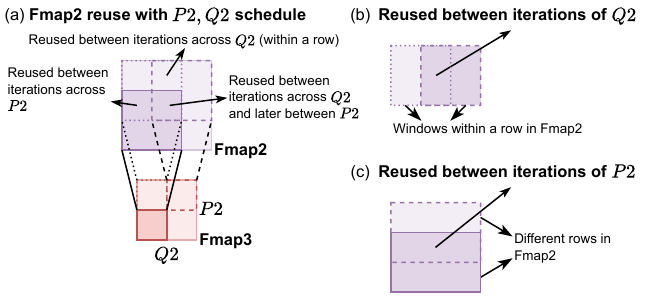}
    \caption{Categorizing data reuse across iterations. (a) Data reuse in Fmap2 tiles with $P2,Q2$ tiling. The reuse can be categorized into two: (b) reuse across iterations of $Q2$ and (c) reuse across iterations of $P2$.}
    \label{fig:design_space:reuse_levels}
\end{figure}

Generally, larger tiles result in more reuse, and we illustrate this fact in terms of our retention choice. Fig.~\ref{fig:design_space:reuse_levels}(a) visualizes the data reuse that results from applying $P2,Q2$ schedule to the tiles in Fig.~\ref{fig:design_space:partitioning_multiple_ranks}. There are two levels of iterations: across the $P2$ rank (rows) and the $Q2$ rank (columns). Note that the iterations are hierarchical: within a single iteration of the $P2$ rank, we iterate over the entire $Q2$ rank. Fig.~\ref{fig:design_space:reuse_levels}(b)-(c) shows the data reuse categorized by iterations. If we retain Fmap2 tiles formed by partitioning $P2$, we will have data reuse between iterations of $P2$ and $Q2$. We have reuse between iterations of $P2$ because the overlap in data tiles in Fig.~\ref{fig:design_space:reuse_levels}(c) can be reused between iterations. We have reuse between iterations of $Q2$ because those iterations happen within a single $P2$ iteration and all the data needed will be retained in the buffer. If we retain Fmap2 tiles formed by partitioning $P2$ and $Q2$, we will only have data reuse between iterations of $Q2$. Between iterations of $P2$, the data in the overlapping region in Fig.~\ref{fig:design_space:reuse_levels}(c) is not guaranteed to still be in the buffer.

Finally, we note that the retained intermediate fmap tile needs to be larger or equal to the intermediate fmap tile produced between layers. Formally, this means that retention choices for intermediate fmaps can only be among the ranks partitioned for the inter-layer tiling. On the other hand, retention choices for non-intermediate-fmap tensors can be made from any of the partitioned ranks, including ranks partitioned in the intra-layer mapping, which we discuss next.

\subsection{Intra-layer Mapping}\label{sec:design_space:intra_layer}
At this point, we have a set of operation and data tiles for each layer. All that is left is to specify the order in which we process the elements inside each tile (\ie, specify the intra-layer mapping). We do not go into the details of the intra-layer mapping choices because they are not the focus of this paper and have been explored extensively in prior work~\cite{timeloop, zigzag, sparseloop, maestro, ruby}. Here, we simply mention that LoopTree supports an extensive set of features from prior work:
\begin{itemize}
    \item intra-layer tiling for each layer independently,
    \item mapping to multiple levels of memory hierarchy~\cite{timeloop, maestro, zigzag, ruby, sparseloop},
    \item whether to process the tiles sequentially or in parallel~\cite{timeloop, maestro, zigzag, ruby, sparseloop},
    \item tiles that are imperfectly factorized~\cite{ruby},
    \item per-tensor retention (also referred to as uneven mapping) mapping~\cite{zigzag}.
\end{itemize}

\section{LoopTree: a Flexible Model for a More Extensive Fused-layer Dataflow Design Space}\label{sec:model}
To explore our \emph{mapspace} (\ie, space of mappings), we need a hardware model for evaluation. Furthermore, we would like this model to be fast and accurate. In this section, we describe a model for our fused-layer dataflow design space. This model leverages patterns in tile shapes and regularity in the hardware behavior when processing identical tile shapes to be able to compute hardware metrics (\eg, latency, energy, buffer occupancy) using mathematical expressions (\ie, the model is \emph{analytical}). Compared to simulators, analytical models tend to be faster at the cost of some fidelity~\cite{sparseloop}.

\begin{figure}[b]
    \centering
    \includegraphics[width=0.42\textwidth]{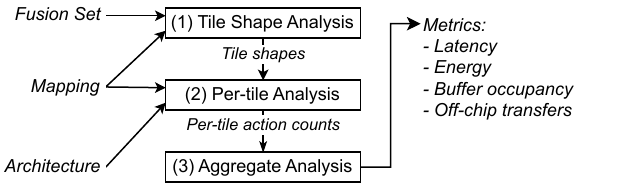}
    \caption{Overview of the LoopTree model.}
    \label{fig:model:overview}
\end{figure}

We briefly overview the analysis steps of the LoopTree model before discussing each step in detail (points match analysis steps in Fig.~\ref{fig:model:overview}).

\begin{enumerate}
    \item Given the mapping, which only specifies the tile shape of the last layer, this step determines the operation and data tile shapes of all other layers in the fusion set.
    \item Given the tile shapes, architecture, and intra-layer mapping, this step counts the number of occurrences of hardware actions (\eg, buffer reads) during the processing of each tile.
    \item Given action counts for each processed tile, this step determines the final metrics (latency, energy, buffer occupancy, and off-chip transfers).
\end{enumerate} 

\subsection{Tile Shape Analysis}\label{sec:model:analyzing_tile_shapes}
\begin{figure}
    \centering
    \includegraphics[width=0.45\textwidth]{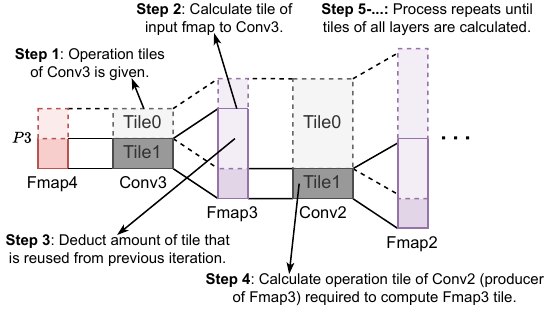}
    \caption{Analyzing tile shape given inter-layer tiling specification. Layers are shown from last to first (in contrast with other figures) to better illustrate the analysis steps.}
    \label{fig:model:tile_analysis}
\end{figure}

The user-defined mapping only specifies the tile shapes of the last layer in the fusion set. The tile shape analysis calculates the tile shapes of all the other layers given the fusion set and the mapping. We describe the analysis steps below (see Fig.~\ref{fig:model:tile_analysis}).
\begin{enumerate}[leftmargin=1.21cm, label=Step {\arabic*)}]
    \item The mapping specifies the operation tiles of the last layer in the fusion set (Conv3 in Fig.~\ref{fig:model:tile_analysis}).
    \item From the operation tile, we can compute the data tile of the input fmap of the last layer (Fmap3 in Fig.~\ref{fig:model:tile_analysis}).
    \item From the data tile of the fmap, we subtract the amount that is retained from previous iterations (in Fig.~\ref{fig:model:tile_analysis}, only a subset of the Fmap3 tile needs to be computed).
    \item The part of the fmap tile that is not retained from previous iterations has to be produced (in Fig.~\ref{fig:model:tile_analysis}, we calculate the operations of Conv2 required to produce it). This may include the recomputation of certain operations.
    \item At this point, we are in the same setup as in step 1, but for an earlier layer (in Fig.~\ref{fig:model:tile_analysis}, Conv2 instead of Conv3). We repeat the analysis until we have calculated the tiles for all the layers in the fusion set.
\end{enumerate}

\begin{figure}[b]
    \centering
    \includegraphics[width=0.475\textwidth]{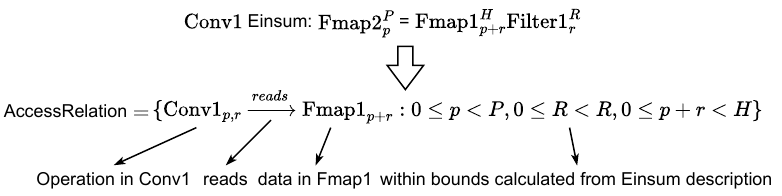}
    \caption{Calculating the data access relation as a polyhedron from the Einsum description. Polyhedrons have efficient implementations for set operations (\eg, intersection, union, etc.).}
    \label{fig:model:polyhedral}
\end{figure}

In the steps we just described, we needed to calculate data tiles from operation tiles and data dependencies (\eg, in step 1) and vice versa (\eg, in step 3). We also needed to perform set operations on operation tiles (\eg, in step 2). We briefly discuss a fast implementation of these calculations. As mentioned at the start of this section, our approach to evaluation is \emph{analytical}. Rather than explicitly constructing operation tiles and the data accesses (\ie, which data each operation accesses), we represent them as a set constrained by equalities and inequalities containing affine expressions (see Fig.~\ref{fig:model:polyhedral}). In other words, we construct polyhedral sets and relations~\cite{loop_optimization}. There are fast methods for performing these set and relation operations~\cite{isl}. While storing sets and relations explicitly requires storage proportional to the size of the DNN we are modeling, which can be enormous, our analytical approach requires storage proportional to the number of tiling steps and the number of ranks in the fusion set. The same is true for the number of operations required in our subroutines. This allows the LoopTree model to be fast (prior work has shown analytical models to be up to $1000 \times$ faster than simulators~\cite{sparseloop}).

\subsection{Per-tile Hardware Action Counts Analysis}\label{sec:model:per_tile_counts}
Given the inter-layer tiles of each layer, architecture, and intra-layer mappings, we now have to analyze the hardware actions (\eg, reads, writes, computes) required when processing each tile. Because this analysis is performed for each layer independently, we can use a similar analysis to layer-by-layer frameworks~\cite{timeloop, maestro, sparseloop, zigzag, ruby} (the analysis in LoopTree is based on Timeloop~\cite{timeloop}). However, we implement this analysis using set and relation operations so that it is compatible with the results of the tile shape analysis\footnote{An alternative is to provide an adapter. However, we hope that the polyhedral implementation will make future extensions easier.}. While this implementation is not trivial, it is not the main focus of this paper. Thus, we refer the readers to the open-source documentation and mention here several notable features of the analysis.
\begin{itemize}
    \item LoopTree analyzes the impact of temporal data reuse on the number of transfers between buffer levels in the memory hierarchy. For example, if the same data is required within a buffer between two iterations, LoopTree detects this reuse opportunity and the data is not refetched from the parent buffer.
    \item We analyze the impact of spatial data reuse (\ie, when the same data is needed by multiple compute units) on buffer reads and data transfers. For example, if two child buffers need to receive the same data at the same time, LoopTree detects the multicast opportunity and counts only a single read to the parent buffer. Furthermore, LoopTree will calculate the number of hops required to send the data from the parent to the child buffers on the network-on-chip (NoC).
\end{itemize}

The result of this analysis is a set of hardware action counts accumulated during the processing of the tiles. Specifically, we have:
\begin{itemize}
    \item Reads and writes to each buffer in the hardware.
    \item The number of network hops to distribute data in each network in the hardware.
    \item The latency required to process all the operations at the compute units.
\end{itemize}

Finally, we note that the processing of tiles with the same shape will have the same behavior and therefore the same hardware counts. Our model detects unique tile shapes and performs the analysis only once for each unique tile shape.

\subsection{Analyzing Final Metrics}
In the final analysis, we use the action counts from the processing of each tile to compute the final metrics.

\subsubsection{Calculating Total Latency}
The latency calculation is divided into two cases: (1) if the tiles are processed sequentially and (2) if the tiles are processed in a pipeline. In the sequential case, latency can be calculated as the sum of the latencies of the processing of each tile.

The pipeline case is more complicated. In Fig.~\ref{fig:model:tile_analysis}, the number of operations in Tile0 and Tile1 Conv2 is different because the reused part of Fmap3 does not have to be recomputed. This makes the number of operations in a tile depend on which iteration the tile belongs to. LoopTree's algorithm for calculating total latency takes this into account.

\begin{figure}[b]
    \centering
    \includegraphics[width=0.49\textwidth]{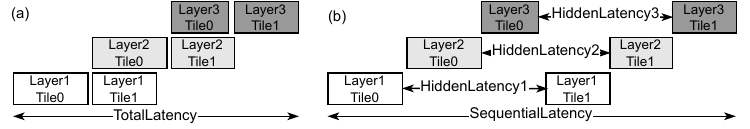}
    \caption{Pipeline latency analysis. (a) Total latency of the processing of three layers in a pipeline. (b) Latency of sequential processing of the layer tiles and the latencies that could be hidden in a pipeline.}
    \label{fig:model:pipeline_latency}
\end{figure}

LoopTree's algorithm for evaluating pipeline latency first arranges the pipeline stages sequentially (\ie, with no pipeline parallelism) (see Fig.~\ref{fig:model:pipeline_latency}(a) and (b)). Overlapping the stages hides some latency. In Fig.~\ref{fig:model:pipeline_latency}(b), they are HiddenLatency1, HiddenLatency2, and HiddenLatency3. The actual hidden latency is the minimum of the three. Then, the total pipeline latency is the sequential latency minus the hidden latency.

Similarly to other analyses in LoopTree, the analysis avoids duplicate work by using a polyhedral representation of tiles.

Finally, the latency we computed so far is only the computation latency. We aggregate read/write counts to each buffer and divide by the bandwidth to get memory latency. We take the larger one of the compute and memory latencies to be the final latency. We note that this analysis assumes that data layout reordering is performed during processing such that the reordering does not increase latency. There are existing state-of-the-art works demonstrating the feasibility of such processing, \eg, the \emph{reorder in reduction} technique introduced by Tong \emph{et al.}~\cite{feather}. Moreover, LoopTree assumes explicit data orchestration using \emph{Buffets}~\cite{buffet} such that pipeline stalls can be assumed to be negligible.

\subsubsection{Calculating Total Energy}
Given the action counts (\ie, read/write counts, the number of computations, network hops, and peer-to-peer transfers) of intra-layer processing (see Section~\ref{sec:model:per_tile_counts}) and inter-layer processing (see Section~\ref{sec:model:analyzing_tile_shapes}), we can calculate the energy consumption of the system by multiplying the counts of each action with the energy per action. The energy per action is generated from the architecture specification using Accelergy~\cite{accelergy}.

\subsubsection{Calculating Buffer Occupancy}
Given a design, we have to make sure that the buffer occupancy does not exceed the buffer capacity. From the tile shape inference, we know the data that has to occupy each buffer. We can then compute the buffer occupancy from the shape of the tiles. The sequential processing of each layer tile may require different amounts of buffer occupancy. LoopTree reports the maximum buffer occupancy.

\subsubsection{Calculating Off-chip Transfers}
As mentioned in Subsection~\ref{sec:model:per_tile_counts}, off-chip transfers (which is a special case of buffer-to-buffer data transfers) are computed per-tile in the last step. The total off-chip transfers are the sum of the per-tile off-chip transfers. LoopTree can also report other buffer-to-buffer transfers.

\section{Validation}\label{sec:validation}

We implement and validate the accuracy of the LoopTree analytical model across prior architectures that cover a wide range of the design choices described in Section~\ref{sec:design_space} and DNNs with varying reuse patterns.

\subsection{Methodology}
We implement LoopTree in C++ as an extension of Timeloop~\cite{timeloop}. We implement the analysis described in Section~\ref{sec:model}, using the ISL library~\cite{isl} to handle set and relation operations. LoopTree uses Accelergy~\cite{accelergy} as the energy estimation back end.

\subsection{Validation Setup}
For validation, we choose designs that exercise a wide range of LoopTree's capabilities. Tab.~\ref{tab:validation:summary} summarizes these designs. We choose five fused-layer dataflow accelerators for convolutional neural networks (CNNs) and transformers. They are chosen because they span different fused-layer dataflow design choices across a range of DNNs. For each design, we model each design executing the same DNNs as the ones they were evaluated on in the publication. Then, based on available information, we compare latency, energy consumption, buffer capacity, and off-chip transfers. 

\begin{table*}
\centering
\caption{High-level summary of validations. DNN layers: convolution (conv), pointwise convolution (pwise), depthwise convolution (dwise), self-attention (sa). Parallelism: sequential (s), parallel (p). Outputs: latency (L), energy (E), capacity (C), off-chip transfers (T).}
\label{tab:validation:summary}
\begin{tabular}{@{}lcccccccccc@{}}
\toprule
Design & DNN type & DNN layers & Partitioned ranks & Retain-recompute & Parallelism & \multicolumn{4}{c}{Output} & \makecell[c]{Max. \\ error \\ \%} \\
& & & & & & L & E & C & T & \\
\midrule
\makecell[l]{DepFin~\cite{depfin}} & \makecell{CNN~\cite{mccnn,fsrcnn}}            & \makecell{conv, dwise, pwise}   & Row, column & Fully retain & s & & $\checkmark$ & $\checkmark$ & $\checkmark$ & 0 \\
Fused-layer CNN~\cite{fusedcnn}    & CNN~\cite{alexnet,vggnet} & conv & Row, column & Fully retain & p & $\checkmark$ & & $\checkmark$ & $\checkmark$ & 1.2 \\
ISAAC~\cite{isaac}                 & CNN~\cite{vggnet}         & conv & Column & Fully retain & p & & $\checkmark$ & $\checkmark$ & & 4 \\
PipeLayer~\cite{pipelayer}         & CNN~\cite{alexnet,vggnet} & conv & Batch & Fully retain & p & $\checkmark$ & & & & 3.3 \\
\makecell[l]{FLAT~\cite{flat}}        & \makecell{Transformers~\cite{bert}} & sa   & \makecell{Heads, tokens, batch} & Fully retain & s & $\checkmark$ &  & $\checkmark$ & $\checkmark$  & 3.4 \\
\bottomrule
\end{tabular}

\end{table*}

\subsection{Validation Results}
Overall, LoopTree shows less than 4\% error. Here, we discuss each validation result in Tab.~\ref{tab:validation:summary} in detail.

\subsubsection{DepFin}
DepFin~\cite{depfin} is a CNN accelerator that partitions $P$ and $Q$ (because the number of layers in the fusion set differs in each validation, we omit the number in the rank name that represents the layer in this section) and processes tiles sequentially with a $P,Q$ schedule. The DNN models are FSRCNN~\cite{fsrcnn}, MC-CNN~\cite{mccnn}, and the two in-house CNN models in the publication~\cite{depfin}. These DNNs contain vanilla convolutions as well as pointwise and depthwise convolutions, which have different reuse patterns. LoopTree results exactly match the energy and off-chip transfer counts as reported in the paper. LoopTree also validates the result that DepFin has enough buffer capacity to retain data given its fusion set, mapping, and architecture.

\subsubsection{Fused-layer CNN}
Fused-layer CNN~\cite{fusedcnn} is a CNN accelerator that partitions $P$ and $Q$ and processes tiles in a pipeline with a $P,Q$ schedule. Because the paper only reports the number of BRAMs used, we create a simulation based on the architecture description for the buffer capacities. Table \ref{tab:fusedcnn_result} shows LoopTree's results, the simulator result, and the result after synthesis on an FPGA reported in the paper. All errors are within 1.2\%. Energy was not reported in~\cite{fusedcnn}.

The slight error in latency is possibly due to implementation details in the synthesized design. The error in off-chip transfers could be caused by LoopTree's assumption of ideal data layout in memory. In practice, off-chip memory block sizes might not match the tile shapes exactly.

\begin{table}[ht]
\centering
\caption{Fused-layer CNN results. Table shows LoopTree, reference, and synthesized result. Reference result is simulated based on pseudocode of the architecture in~\cite{fusedcnn}. \figtakeaway{LoopTree results are within 1.2\%}.}
\begin{tabular}{@{}lccc@{}}
\toprule
    Metric & LoopTree & Ref. & Synth. \\
\midrule
    Latency (kcycles) & 422 &  & 427 \\
    WBuf capacity (KB) & 167 & 167 &  \\
    IOBuf capacity (KB) & 44 & 44 &  \\
    TBuf capacity (KB) & 6 & 6 & \\
    Off-chip transfers (KB) & 611 & 611 & 688 \\

\bottomrule
\end{tabular}
\label{tab:fusedcnn_result}
\end{table}

\subsubsection{ISAAC}
ISAAC~\cite{isaac} is a CNN accelerator that partitions $Q$ and processes tiles in a pipeline. We model their dataflow in LoopTree using the same assumption of balanced throughput. Table \ref{tab:isaac_result} shows the buffer capacities calculated by LoopTree and the reported values. Furthermore, we model their energy consumption in LoopTree. LoopTree recreates their reported energy efficiency with at most 4\% error. The difference could be attributed to a slight mismatch between LoopTree's action-based modeling and ISAAC's power-throughput model.

\begin{table}[b]
\centering
\caption{ISAAC buffer capacity requirement. Reference is the results in~\cite{isaac}. \figtakeaway{LoopTree matches reference results.}}
\begin{tabular}{@{}lcc@{}}
\toprule
    DNN & LoopTree (KB) & Ref. (KB) \\
\midrule
    VGG-1-conv1 & 1.96 & 1.96 \\
    VGG-1-conv2 & 21 & 21 \\
    VGG-1-conv3 & 21 & 21 \\
    VGG-1-conv5 & 21 & 21 \\
\bottomrule
\end{tabular}
\label{tab:isaac_result}
\end{table}

\subsubsection{PipeLayer}
PipeLayer~\cite{pipelayer} is a CNN accelerator that partitions $B$ and processes tiles sequentially or in a pipeline. PipeLayer shows speedup when processing tiles in a pipeline. We model both their sequential and pipelined dataflow in LoopTree. LoopTree replicates the latency results in~\cite{pipelayer} with 3.3\% error (see Tab. \ref{tab:pipelayer_result}). The difference could be due to unmodeled aspects of the ReRAM arrays (\eg, latency from loading weights).

\begin{table}[h]
\centering
\caption{PipeLayer speedup due to pipelining. Reference from~\cite{pipelayer}. \figtakeaway{LoopTree results are within 3.3\% of reference.}}
\begin{tabular}{@{}lcc@{}}
\toprule
    DNN & LoopTree & Ref. \\
\midrule
    AlexNet & 4.8 & 4.8 \\
    VGG-A   & 7.9 & 8.0 \\
    MNIST-A & 2.0 & 2.0 \\
    MNIST-B & 2.9 & 3.0 \\
\bottomrule
\end{tabular}
\label{tab:pipelayer_result}
\end{table}

\subsubsection{FLAT}
\begin{figure}[b]
    \centering
    \begin{subfigure}[b]{0.24\textwidth}
        \includegraphics[width=\textwidth]{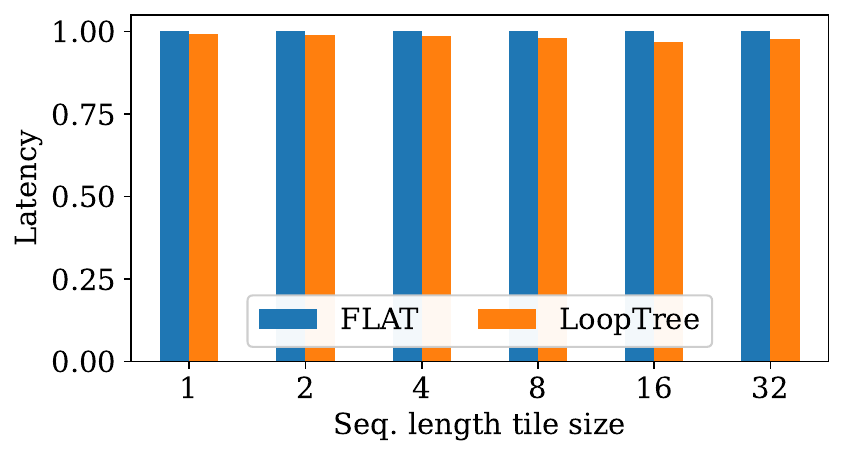}
    \end{subfigure}
    \begin{subfigure}[b]{0.24\textwidth}
        \includegraphics[width=\textwidth]{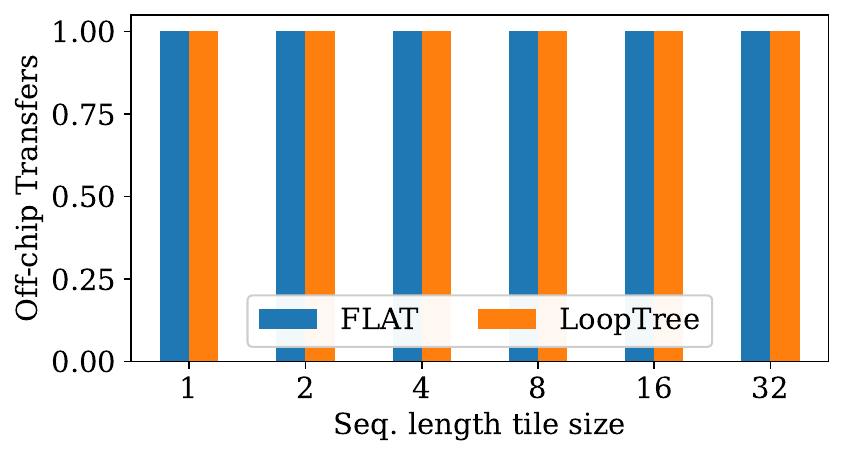}
    \end{subfigure}
    \caption{Normalized (a) latency and (b) off-chip transfers generated by the FLAT~\cite{flat} simulator and LoopTree (this work). \figtakeaway{Results differ by at most 3.4\%.}}
    \label{fig:validation:flat}
\end{figure}

FLAT~\cite{flat} is an accelerator for transformers~\cite{attention} that partitions $B$, $H$, and $M$ and processes tiles sequentially with a $B,H,M$ schedule. We modeled the FLAT dataflow with different tile shapes using LoopTree. Fig.~\ref{fig:validation:flat} shows the normalized results. In all experiments, LoopTree results differ by at most 3.4\%. The most divergent results are latency. The small difference stems from aspects in the FLAT simulator that LoopTree does not model (\eg, latency from loading weights and systolic array startup).

\section{Case Studies}\label{sec:case_study}
We discuss a series of case studies that illustrate the trade-off between on-chip buffer capacity, recomputation, and off-chip transfers in fused-layer dataflow design. Specifically, we evaluate the impact of design aspects in Tab.~\ref{tab:introduction:comparison} (formalized in Section~\ref{sec:design_space}) and how they interact.

Tab.~\ref{tab:case_study:overview} summarizes the setup for each case study. Each case study investigates the impact of design space choices by setting those choices as independent variables and searching for other choices that lead to the optimal in a certain metric. Specifically, in Section~\ref{sec:case_study:tiling_choice}, we evaluate the impact of (inter-layer) partitioned ranks and tile processing schedule (or, ``schedule" for short) without recomputation. In Section~\ref{sec:case_study:further_tiling}, we allow recomputation and evaluate the impact of partitioned ranks and schedule choices on the trade-off between recomputation and on-chip buffer capacity. In Section~\ref{sec:case_study:per_fmap_choice}, we look at the impact of being able to make recomputation choices per-intermediate-fmap. Then, in Section~\ref{sec:case_study:per_tensor_retain}, we show how per-tensor retention leads to a smaller required on-chip buffer capacity. Finally, in Section~\ref{sec:case_study:further_tiling}, we show how LoopTree can be used to evaluate the overall impact of tiled fusion.

\begin{table*}
\centering
\caption{Overview of case study setup. The ``Case study" column refers to the subsection that discusses the case study.}
\label{tab:case_study:overview}
\begin{tabular}{@{}cccccc@{}}
\toprule
\makecell[c]{Case study} & Partitioned Rank & Tile Shape & Schedule & \makecell[c]{Retention \\ (for non-intermediate tensors)} & \makecell[c]{Retention-recomputation \\ (for intermediate fmaps)}  \\
\midrule
B & Independent & Searched & Independent & Searched & Searched s.t. no recomputation \\
C & Independent & Searched & Independent & Searched & Searched \\
D & Searched & Searched & Searched & Independent & Searched \\
E & Searched & Searched & Searched & Searched & Independent \\
F & Searched & Searched & Searched & Searched & Searched s.t. no recomputation \\
\bottomrule
\end{tabular}
\end{table*}

\subsection{Introducing the Fusion Sets}\label{sec:case_study:fusion_sets}
Before discussing the case studies, we discuss the three fusion sets we will use. Tab.~\ref{tab:case_study:fused_sets} shows the Einsums of the fusion sets, which tells us the distinctive features of each fusion set:
\begin{enumerate}
    \item Two 2D convolutions (conv) layers. These are the 2D convolutional layers we have discussed so far.
    \item Three layers consisting of pointwise (pwise), depthwise (dwise), and pointwise (pwise) convolutions. A notable difference from conv layers is that pwise layers do not have convolutional reuse (\eg, Fmap3 in Tab.~\ref{tab:case_study:fused_sets} have $p3$ instead of $p3+r3$ index) and dwise layers do not have channel reuse (in Tab.~\ref{tab:case_study:fused_sets}, the $M2$ rank is shared by Fmap2, Fmap3, and Filter2).
    \item Two fully connected layers (fc) modeled after those in transformers~\cite{bert, xlm, attention}. This fusion set only has full reuse and no reuse; there is no convolutional reuse.
\end{enumerate}

\renewcommand{\arraystretch}{1}
\begin{table*}[h]
\centering
\caption{Layers comprising the fusion sets we will use to evaluate fused-layer dataflows. The rank shapes column shows which ranks have equal values. Bolded text denotes rank values which we will vary throughout the experiment to get fusion sets of different shapes.}
\begin{tabular}{@{}lccc@{}}
\toprule
    Fusion set & Einsums & Fusion set shapes & Modeled after \\
\midrule
    conv+conv
    & 
    $\begin{aligned}
        \text{Fmap2}_{m1,p1,q1}^{M1,P1,Q1} &= \text{Fmap1}_{c1,p1+r1,q1+s1}^{M1,P1,Q1}\text{Filter1}_{c1,m1,r1,s1}^{C1,M1,R1=3,S1=3} \\
        \text{Fmap3}_{m2,p2,q2}^{M2,P2,Q2} &= \text{Fmap2}_{c2,p2+r2,q2+s2}^{C2,P2,Q2}\text{Filter2}_{c2,m2,r2,s2}^{C2,M2,R2,S2} \\
    \end{aligned}$
    &
    \makecell{\textbf{Rows}: $P1=Q1=P2=Q2$ \\ \textbf{Channel}: $C1=M1=C2=M2$}
    &
    \makecell{ResNet \\ blocks~\cite{resnet}}
    \\ \midrule

    pwise+dwise+pwise
    &
    $\begin{aligned}
        \text{Fmap2}_{m1,p1,q1}^{M1,P1,Q1} &= \text{Fmap1}_{c1,p1,q1}^{C1,P1,Q1}\text{Filter1}_{c1,m1}^{C1,M1} \\
        \text{Fmap3}_{m2,p2,q2}^{M2,P2,Q2} &= \text{Fmap2}_{m2,p2+r2,q2+s2}^{M2,P2,Q2}\text{Filter2}_{m2,r2,s2}^{M2,R2=3,S2=3} \\
        \text{Fmap4}_{m3,p3,q3}^{M3,P3,Q3} &= \text{Fmap3}_{c3,p3,q3}^{C3,P3,Q3}\text{Filter2}_{c3,m3}^{C3,M3} \\
    \end{aligned}$
    &
    \makecell{\textbf{Rows}: $P1=Q1=P2=Q2$ \\ \textbf{Channel}: $C1=M3$ \\ $C1=\frac{M1}{6}=\frac{M2}{6}=\frac{C3}{6}$ }
    &
   \makecell{MobileNetv2 \\ blocks~\cite{mobilenetv2}}
    \\ \midrule

    fc+fc
    &
    $\begin{aligned}
        \text{Fmap2}_{m1,e1}^{M1,E1} &= \text{Fmap1}_{m1,d1}^{M1,D1}\text{Filter1}_{d1,e1}^{D1,E1} \\
        \text{Fmap3}_{m2,e2}^{M2,E2} &= \text{Fmap2}_{m2,d2}^{M2,D2}\text{Filter2}_{m2,e2}^{M2,E2} \\
    \end{aligned}$
    &
    \makecell{\textbf{Tokens}: $M1=M2$ \\ \textbf{Emb. dims.}: $E1=D2$ \\ $D1=E2=1024$}
    &
    \makecell{Transformer \\ feed-forward \\ blocks~\cite{attention}}
    \\
\bottomrule
\end{tabular}

\label{tab:case_study:fused_sets}
\end{table*}
\renewcommand{\arraystretch}{1}

In the Einsums, we constrain the fusion set shapes to match common shapes in recent DNNs. Tab.~\ref{tab:case_study:fused_sets} shows the shape of the layers in the fusion sets. (For conciseness, we will refer to the ``shapes of layers in a fusion set" collectively as the \emph{fusion set shape}.) The bolded rank names are variables we will change to vary the fusion set shapes.

\subsection{Impact of Partitioned Ranks and Schedule Choices}\label{sec:case_study:tiling_choice}
\begin{figure*}[h]
    \centering
    \includegraphics[width=0.9\textwidth]{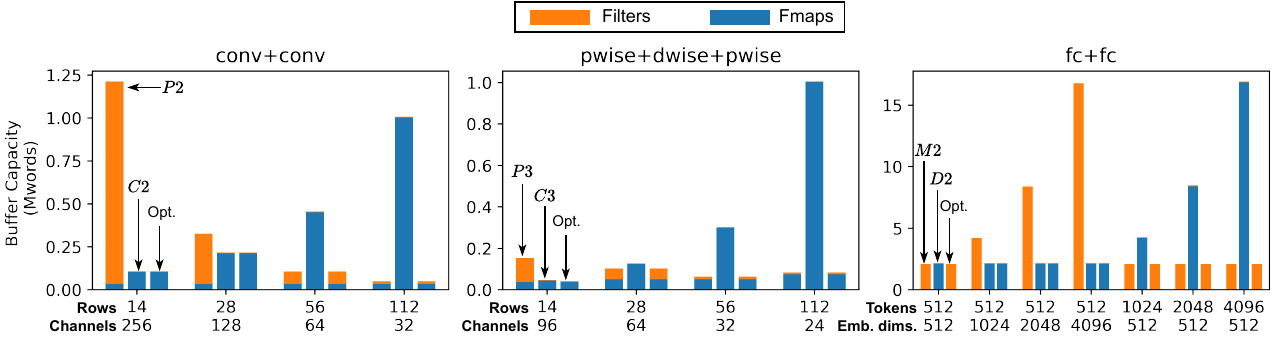}
    \caption{Buffer capacity required for alg. min. off-chip transfers using different partitioned rank and schedule choices. Subplots: different fusion sets. Groups of bars: different fusion set shapes (see Tab.~\ref{tab:case_study:fused_sets} col. 3) Bars: different partitioned ranks and schedules. We show three bars: the optimal (opt.) choice and two other choices for comparison. Partitioned ranks and schedule choices have a significant impact on buffer capacity across different types of fusion sets and fusion set shapes}
    \label{fig:case_study:evaluating_tiling}
\end{figure*}

In this case study, we evaluate the impact of partitioned ranks and schedule choices on the required on-chip buffer capacity to achieve algorithmic minimum off-chip transfers (\ie, the minimum achievable off-chip transfers assuming infinite on-chip buffer capacity) without recomputation. To simplify the following discussion, we refer to only the schedule choice and imply the partitioned ranks. For example, a $P2,C2$ schedule implies that we create tiles by partitioning $P2$ and $C2$, and the tiles are processed with a $P2,C2$ schedule. Fig.~\ref{fig:case_study:evaluating_tiling}, shows the on-chip buffer capacity required by the optimal partitioned ranks and schedule choices (which change with fusion set shape) and two other choices for comparison.

Fig.~\ref{fig:case_study:evaluating_tiling} shows that partitioned ranks and schedule choices have a significant impact on the required on-chip buffer capacity (\eg, the capacity required by a $P2$ and $C2$ schedule may differ by up to $10\times$). The reason is that the partitioned ranks and schedule determine which tensor needs to be stored on-chip. For example, with the $P2$ schedule in conv+conv, we fully reuse Filter1 and Filter2 because they are required to process every operation tile. Thus, if we do not want to refetch Filter1 and Filter2 multiple times from off-chip buffers, we must keep those tensors on-chip. Then, in fusion sets where Filter1 and Filter2 are large (\eg, when there are many channels), Filter1 and Filter2 significantly increase the required on-chip buffer capacity.

As a corollary of the above, the optimal choice tends to correspond with the largest rank. For example, when there are many rows (\ie, $P2$ is large), the $P2$ schedule results in a smaller required on-chip buffer capacity. This is because, in fusion sets where the $P2$ rank is large compared to other ranks, tensors with the $P2$ rank tend to be larger than other tensors (\eg, the size of Fmap1 is proportional to $P2$, but Filter1 is not). Thus, partitioning $P2$ means we are tiling the larger tensors and fully reusing the smaller ones. Because we have to store the entirety of the fully reused tensors, this leads to a smaller on-chip buffer capacity.

The pwise+dwise+pwise fusion set in Fig.~\ref{fig:case_study:evaluating_tiling} shows a similar trend in buffer capacities and breakdowns as the conv+conv fusion set. However, the total filter size is generally smaller, which is a feature of the MobileNet design~\cite{mobilenetv2}. As a result, the $P3$ schedule, which is better when filters are smaller than fmaps, is the optimal choice for more fusion set shapes.

The reasoning can also be applied to fully-connected layers. For example, using the $M2$ schedule results in fully reusing (and thus retaining the entirety of) the filters. On the other hand, using the $D2$ schedule results in retaining the entirety of Fmap1 and Fmap3. Thus, we see the pattern in Fig.~\ref{fig:case_study:evaluating_tiling}.

\textbf{Takeaway 1: the partitioned ranks and schedule that results in the smallest required on-chip buffer capacity is often the one that reuses the smallest tensors.}

\subsection{Impact of Partitioned Ranks and Schedule Choices with Recomputation}\label{sec:case_study:further_tiling}
\begin{figure}
    \centering
    \includegraphics[width=0.495\textwidth]{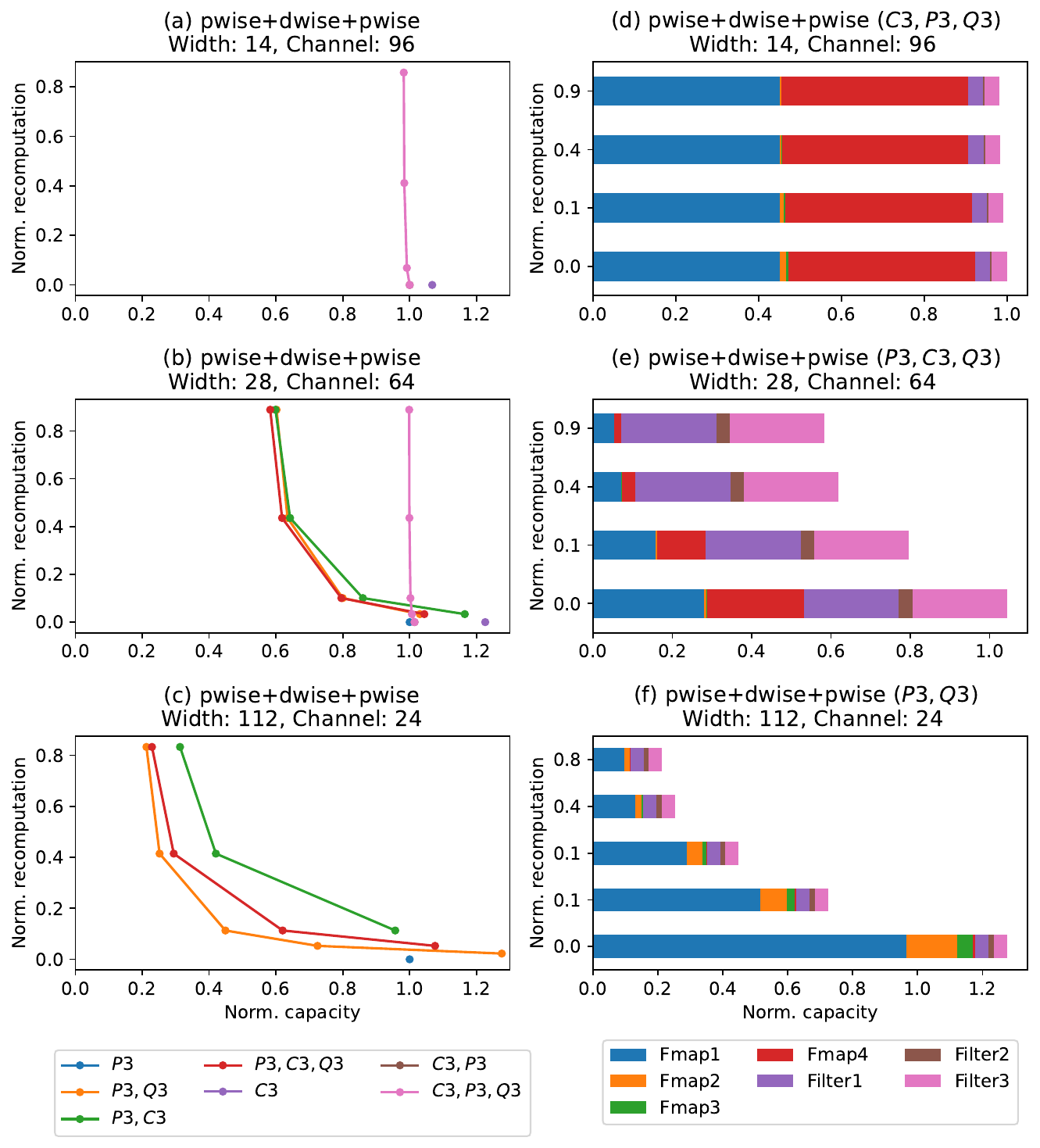}
    \caption{(a)-(c) Normalized recomputation against the normalized capacity for different partitioned ranks and schedule (colors) and fusion set shapes (subgraphs). The optimal partitioned ranks and schedule choice vary by fusion set shape. (d)-(f) Breakdown of buffer capacity usage by tensors for the partitioned ranks and schedule choice in parentheses.}
    \label{fig:case_study:pwise_dwise_pwise}
\end{figure}

In this case study, we evaluate the impact of partitioned ranks and schedule choices on the required on-chip buffer capacity to achieve algorithmic minimum off-chip transfers with recomputation. We compare the Pareto front of required on-chip buffer capacity and recomputation (\ie, the set of mappings that achieve the fewest recomputations and smallest required on-chip buffers) for different partitioned ranks and schedule choices. Fig.~\ref{fig:case_study:pwise_dwise_pwise} shows the Pareto front for pwise+dwise+pwise. We make three observations.

First, recomputation changes which partitioned ranks and schedule result in a smaller required on-chip buffer capacity. For example, Fig.~\ref{fig:case_study:pwise_dwise_pwise}(b) and (c) show that the partitioned ranks and schedule that results in the smallest required on-chip buffer capacity without recomputation is $P3$, which is different with recomputation ($P3,C3,Q3$ in Fig.~\ref{fig:case_study:pwise_dwise_pwise}(b) and $P3,Q3$ in Fig.~\ref{fig:case_study:pwise_dwise_pwise}(c)). Thus, retention-recomputation, partitioned ranks, and schedule choices need to be explored together.

Second, the partitioned ranks and schedule that results in the most efficient recomputation and capacity trade-off differs for different fusion set shapes (\eg, it is $P3,C3,Q3$ in Fig.~\ref{fig:case_study:pwise_dwise_pwise}(b) and $P3,Q3$ in Fig.~\ref{fig:case_study:pwise_dwise_pwise}(c)). Thus, it is important to search from an extensive set of partitioned rank choices because no particular choice results in the smallest required on-chip buffer capacity at a given recomputation amount for all fusion set shapes.

The reason behind the second observation is similar to the one discussed in the last subsection. The partitioned ranks and schedule choice determine which tensors are fully reused and thus fully retained. In Fig.~\ref{fig:case_study:pwise_dwise_pwise}(a), there are many channels, but the width is small, thus the filters are larger than fmaps. Any of the partitioned ranks and schedule that starts with $P3$ needs to fully retain the filters, which significantly increases the required on-chip buffer capacity. Comparing Fig.~\ref{fig:case_study:pwise_dwise_pwise}(a), (b), and (c), the trend reverses as fmaps become larger than filters.

Third, note that the slope of the Pareto frontier differs for different partitioned ranks and schedules. When the slope is steep (\eg, $C3,P3,Q3$ in Fig.~\ref{fig:case_study:pwise_dwise_pwise}(a)), more recomputation does not lead to significant required on-chip buffer capacity reduction. The breakdown of the capacity usage (see Fig.~\ref{fig:case_study:pwise_dwise_pwise}(d)) explains the steep slope: with the $C3,P3,Q3$ schedule, Fmap1 and Fmap4 need to be fully retained. We can reduce buffer capacity by retaining smaller tiles of Fmap2 and Fmap3 at the cost of more recomputation. However, because the Fmap2 and Fmap3 tiles are only small portions of the required capacity, the reduction in required capacity is insignificant. Compare this with $P3,Q3$ in Fig.~\ref{fig:case_study:pwise_dwise_pwise}(c) and (f). With $P3,Q3$, the filters need to be fully retained. However, the fmaps still take up the majority of the on-chip buffer capacity. Thus, retaining smaller tiles of the fmaps results in significant capacity reduction. 

Finally, we note that the fc+fc fusion set does not have retention-recomputation choices because all partitioned ranks and schedule choices for fc+fc result in intermediate fmap tiles that do not overlap.

\textbf{Takeaway 2: retention-recomputation, partitioned ranks, and schedule choices need to be explored together, and it is important to search from an extensive set of partitioned rank choices because no particular choice results in the smallest buffer capacity for all fusion set shapes.}

\subsection{Impact of Per-tensor Retention Choices}\label{sec:case_study:per_tensor_retain}

\begin{figure}
    \centering
    \includegraphics[width=0.495\textwidth]{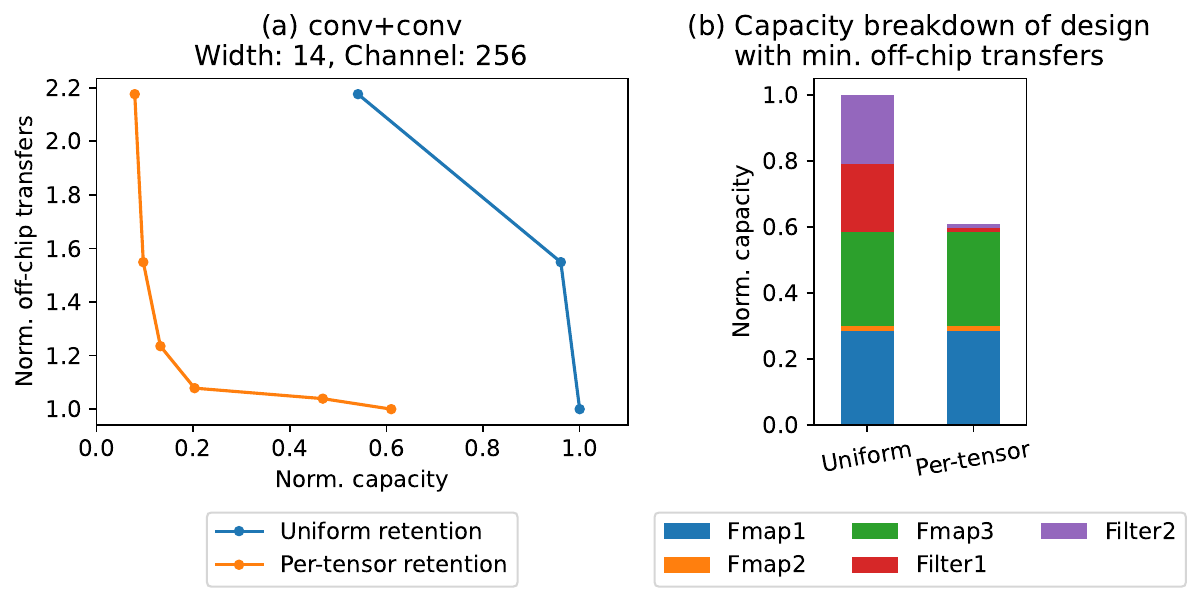}
    \caption{(a) Normalized off-chip transfers against normalized buffer capacity with uniform and per-tensor retention. Using per-tensor retention can reduce buffer capacity by up to 40\%. (b) Capacity breakdown of the design with min. off-chip transfers for uniform and per-tensor retention.}
    \label{fig:case_study:per_tensor_retain}
\end{figure}

We evaluate the impact of having per-tensor retention choices compared to uniform retention choices for the conv+conv fusion set (other fusion sets show similar results). In this evaluation, we do not consider recomputation. Because the LoopTree mapspace allows per-tensor retention, we explore the default mapspace to evaluate mappings with per-tensor retention. To evaluate mappings with uniform retention, we constrain the mapspace such that the same retention choice is made for all tensors. Fig.~\ref{fig:case_study:per_tensor_retain}(a) shows normalized off-chip transfers against normalized on-chip buffer capacity for the conv+conv fusion set with per-tensor retain and uniform retain.

Fig.~\ref{fig:case_study:per_tensor_retain}(a) shows that per-tensor retention can reduce required on-chip buffer capacity significantly (up to $9\times$). To illustrate why, we pick the per-tensor and uniform retention mappings with the fewest off-chip transfers (the lowest point of each curve) and show the capacity usage breakdown in Fig.~\ref{fig:case_study:per_tensor_retain}(b). Filter1 and Filter2 require much smaller on-chip buffer capacity with per-tensor rather than with uniform retention. For a given schedule, different tensors have different minimum tile shapes that need to be retained to avoid refetches. Making per-tensor retention choices allows us to match these minimum tile shapes. Here, the uniform retention choice retains larger filter tiles than necessary to avoid refetches due to its constraints.

\textbf{Takeaway 3: Per-tensor retention choices result in a smaller required on-chip buffer capacity because we can adapt the retention choice to the reuse pattern of each tensor}.

\subsection{Impact of Per-tensor Retention-recomputation Choices}\label{sec:case_study:per_fmap_choice}
\begin{figure}
    \centering
    \includegraphics[width=0.495\textwidth]{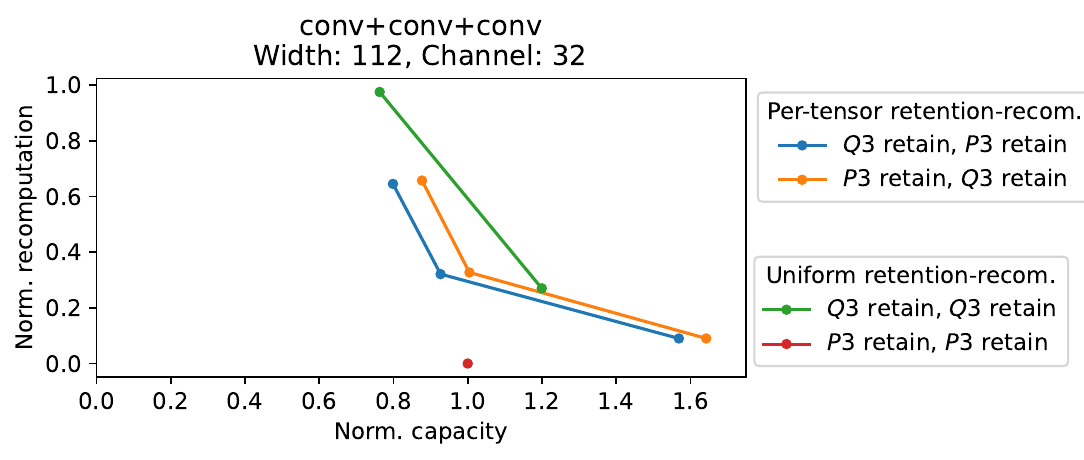}
    \caption{Normalized buffer capacity and normalized recomputation Pareto curves for $P2,Q2$ for different retain-recompute choices. The legend lists retain-recompute choices for Fmap2 and Fmap3 respectively. \figtakeaway{Mixing retain-recompute choices for different fmaps leads to better tradeoffs than a uniform retain-recompute choice}.}
    \label{fig:case_study:mixing_retain_recompute}
\end{figure}

When there are multiple intermediate fmaps, LoopTree allows us to make retention-recomputation choices per tensor. Here, we evaluate the impact of per-tensor retention-recomputation choices on a fusion set of three convolutional layers (conv+conv+conv)\footnote{This fusion set is omitted in Tab.~\ref{tab:case_study:fused_sets} because it is used only in this case study. We use three conv layers such that we have two intermediate fmaps that have retention-recomputation choices. We do not use pwise+dwise+pwise because Fmap3 does not have retention-recomputation choices.} for the $P3,Q3$ schedule. There are two intermediate fmaps in the fusion set (Fmap2 and Fmap3) and for each intermediate fmap, we can either choose to retain across $P3$ or $Q3$. Without support for per-tensor choices, we can only make the same choice for all the tensors (\ie, uniform choices). By being able to make per-tensor retention-recomputation choices, we can also make different choices in addition to uniform choices. Fig.~\ref{fig:case_study:mixing_retain_recompute} shows that per-tensor choices result in a smaller required on-chip buffer capacity for the same recomputation amount. Furthermore, recomputing Fmap2 and retaining Fmap3 leads to a smaller required on-chip buffer capacity than recomputing Fmap3 and retaining Fmap2.

To understand the result, note that recomputing Fmap3 requires more activations in Fmap2 to be retained as inputs. If we are retaining Fmap2, then a larger on-chip buffer is required to retain Fmap2. If we are recomputing Fmap2, then the amount of Fmap2 recomputation increases compared to if we retained Fmap3 (\ie, whether we choose to retain or recompute Fmap3 impacts the number of recomputations of Fmap2). Thus, recomputing later layers increases the required on-chip buffer capacity or recomputations in earlier layers. Prior work referred to this as a compounding effect in recomputation~\cite{fusedcnn}.

\textbf{Takeaway 4: per-tensor retention-recomputation choices can reduce the amount of recomputation by limiting the compounding of recomputations.}

\subsection{Overall Impact of Tiled Fusion}\label{sec:case_study:fuse_or_not}
\begin{figure}[b]
    \centering
    \includegraphics[width=0.495\textwidth]{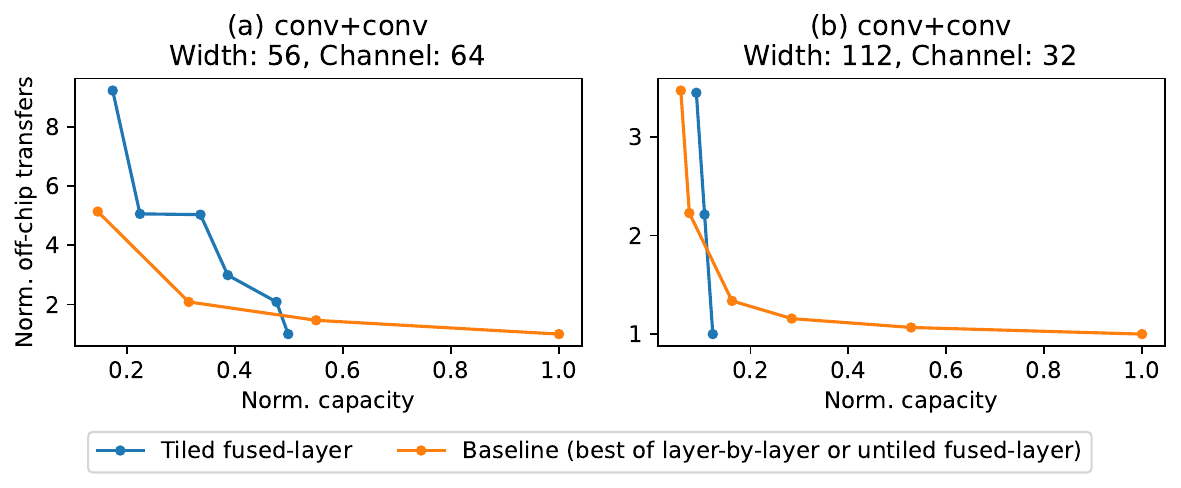}
    \caption{Buffer capacity and off-chip transfers tradeoff Pareto curves for tiled fused-layer and baseline without recomputation. \figtakeaway{At capacity lower than required for minimum algorithmic off-chip transfers, the baseline is often better than fused-layer.}}
    \label{fig:case_study:lbl+pq}
\end{figure}

In this case study, we compare tiled fused-layer designs against a baseline that picks the best of either processes layer-by-layer or retains entire intermediate fmaps (\ie, untiled fusion). We take the conv+conv fusion set with shapes where $P2,Q2$ is the optimal choice. Then, we evaluate the required on-chip buffer capacity required for different amounts of off-chip transfers without recomputation. Fig.~\ref{fig:case_study:lbl+pq} shows the results.

In Fig.~\ref{fig:case_study:lbl+pq}, we can see that the number of off-chip transfers rises faster for the tiled fused-layer dataflow as available on-chip buffer capacity is reduced. As a result, the baseline dataflow is more efficient when on-chip buffers are small.

To see why, we make two observations:
\begin{itemize}
    \item Intermediate fmap activations are reused between layers twice: one read and write saved for every activation.
    \item Intra-layer reuse is often more abundant. For example, in the conv+conv fusion set with 64 channels, each activation in the fmap is read by $3\times3\times 64=576$ operations (activations near the borders are reused fewer times, but these are few).

\end{itemize}
Thus, given limited capacity, it is more efficient to exploit the more abundant intra-layer reuse.

On the other hand, achieving algorithmically minimum accesses using tiled fused-layer dataflow requires a significantly smaller capacity, compared to the baseline which has to retain the entire intermediate fmap to achieve minimum accesses. This result shows the advantage of inter-layer tiling in achieving inter-layer reuse efficiently.

Of course, there exist factors that may cause fused-layer dataflows to be more efficient than layer-by-layer even if the available on-chip capacity is low: certain workloads may have few intra-layer reuse opportunities (\eg, in elementwise operations) or if compute units are especially efficient, which may make recomputation cheaper. As we have demonstrated, LoopTree can be used to explore these trade-offs to find efficient designs.

\textbf{Takeaway 5: at buffer capacities much lower than required for algorithmic minimum off-chip transfers, fused-layer dataflows are often less efficient than layer-by-layer dataflows.}

\section{Related Works}\label{sec:related_works}
\subsection{Fused-layer Dataflow Accelerators}
Prior work has proposed fused-layer dataflow accelerators (under various names)~\cite{atomic, tangram, flat, fusedcnn, depfin, isaac, pipelayer}. These works have shown that fused-layer dataflow accelerators can have lower latency, lower energy consumption, or higher scalability compared to layer-by-layer dataflow accelerators. This makes the fused-layer dataflow accelerator design space interesting to explore. However, although these works often describe a hardware performance model and an algorithm to search for an optimal mapping, each model and mapping search approach are specific to the proposed accelerator and the target workload. Thus, a more general model that supports a wide fused-layer dataflow accelerator design space and various workloads is needed.

\subsection{Fused-layer Dataflow Accelerator Exploration Framework}
Fused-layer dataflow exploration frameworks (\eg,~\cite{defines, set, tileflow, optimus, convfusion} often support a range of fused-layer dataflow accelerators (instead of a specific one) and workloads. However, as discussed in Sections~\ref{sec:introduction} and~\ref{sec:background}, prior frameworks have limitations in certain features of their design spaces. This work addresses the aforementioned limitations. Furthermore, although each framework may have only a subset of the limitations, our case studies in Section~\ref{sec:case_study} have shown that exploring these features in combination results in more efficient accelerator designs.

Prior work has also proposed methods for determining optimal fusion sets~\cite{optimus, set, convfusion, atomic}. For example, Optimus~\cite{optimus} is a method based on dynamic programming, SET~\cite{set} is based on simulated annealing, ConvFusion~\cite{convfusion} is based on Pareto-filtering. Orthogonally, LoopTree is a model to find the optimal design choices for a fusion set. Thus, LoopTree can be used in conjunction with any of these methods.

\subsection{Search Algorithms for Design Space Exploration}
Prior work has explored using various search algorithms to explore the mapspace~\cite{cosa, mindmappings, gamma_mapper, convfusion, optimus, set, atomic}. Although not necessarily exploring the mapspace of fused-layer accelerators, many of these search algorithms can be adapted to search the LoopTree mapspace using LoopTree as the model, \eg, dynamic programming~\cite{optimus}, genetic algorithms~\cite{gamma_mapper}, or differentiable surrogate in~\cite{mindmappings}.

\section{Conclusion}
In this paper, we described a fused-layer mapspace with more extensive tiling, recomputation, and retention choices than prior work. We also described and validated an analytical hardware model that can evaluate latency, energy, and required buffer capacity given a mapping and fusion set. Using this model, we explore the fused-layer mapspace for fusion sets in CNNs and transformers, revealing insights into fused-layer dataflow accelerator design that can only be explored with a model supporting a more extensive fused-layer dataflow design space.

\section*{Acknowledgements}
This work was funded in part by the MIT AI Hardware Program.





\ifCLASSOPTIONcaptionsoff
  \newpage
\fi



%
\bibliography{refs}
\bibliographystyle{IEEEtran}


\end{document}